# Performance Analysis of Adaptive Noise Cancellation for Speech Signal

Pratibha Balaji, Shruthi Narayan, Durga Sraddha, Bharath K P, Karthik R, Rajesh Kumar Muthu
School of Electronics Engineering
VIT University, Vellore, India.
balajipratibha@gmail.com,shruthi.narayan98@gmail.com,sraddhadurga@gmail.com,bharathkp25@gmail.com,tkgravikarthik@gmail.com,
mrajeshkumar@vit.ac.in

*Abstract*-This paper gives a broader insight on the application of adaptive filter in noise cancellation during various processes where signal is transmitted. Adaptive filtering techniques like RLS, LMS and normalized LMS are used to filter the input signal using the concept of negative feedback to predict its nature and remove it effectively from the input. In this paper a comparative study between the effectiveness of RLS, LMS and normalized LMS is done based on parameters like SNR (Signal to Noise ratio), MSE (Mean squared error) and cross correlation. Implementation and analysis of the filters are done by taking different step sizes on different orders of the filters.

*Keywords—Adaptive filter, RLS, LMS, NLMS, SNR, Correlation Coefficient, MSE*

## I. INTRODUCTION

Many different applications have been invented and introduced in the communication field [1]. One of the most basic parameters needed in a communication is the transmitting signal. Noise, which is an unwanted disturbance, is added through the channel to our speech signal. There are different types of noise present, like additive white Gaussian noise, shot noise, thermal noise, random noise amongst many others, add themselves onto the signal and create problems by corrupting the original signal. They are many methods available to remove this interference from the original signal. The method implied here to de-noise the original message signal is Adaptive Noise Cancellation. Adaptive Noise Cancellation or Active Noise Control (ANC) is a method in which a reference noise, which is inputted, is adaptively subtracted from the original noise signal [2]. The advantage of ANC lies in the fact that the noise cancellation can be done to such an extent that no other digital signal processing tool can attain. Suppose an input signal X, which has N noise already added to it, is transmitted and received.

Upon filtering using adaptive techniques, an estimation of noise signal 'n' obtained from the inputted reference noise $N_0$. This estimated signal is subtracted from the input. The final output after filtering is X+ (N-n) depicted as d (n) as seen in the Fig.1 [3]. There are various kinds of adaptive noise control algorithms present, among which we would be employing LMS, NLMS and RLS algorithms to cancel the noise and compare their abilities in de-noising the signals [4].

Shubra et al., has stated through her research that though RLS provides better results, LMS is a better option as it has simpler calculations and stability [1]. Haykin describes adaptive noise cancellation and the different techniques associated with it in a cohesive manner for better understanding [2]. Proakis, has mentioned the different noise filtering methods available in signal processing[3] . Mendiratta et al., has discussed in detail about Adaptive noise cancelling for audio signals using least mean square algorithm with successful results [4]. Thenua et al., describes the performance analysis and has provided with appropriate simulation results and has stated that NLMS is a better choice due to its less computational complexity and better noise reduction [5]. Pandey et al., conducted an effective comparison study between RLS and LMS algorithms and concluded by proving that RLS is a better technique than LMS as it provided better results. [6]. Borisagar et al., has simulated and provided a comparative study on RLS and LMS algorithms[7].Panda et al., have done a thorough analysis on the real time noise cancellation using adaptive filters using the most commonly used technique LMS[8]. Gupta et al, has done a comparative study on the algorithm techniques like LMS, NLMS and Unbiased and Normalized Adaptive Noise Reduction (UNANR) using parameters like SNR and PSNR [9]. Sharma et al., has conducted a thorough analysis on modified normalized least mean square algorithm [10]. Yang Liu et al., has efficiently filtered out the noise from the speech signal by applying NLMS technique [13]. Divya et al have proved with results that NLMS is a better adaptive filter than that of LMS filter and it also has a better convergence rate. They have stated through their research that larger the step size, larger should be the filter length to yield definite results [14].

Out of the three methods employed in this paper, after an elaborate and meticulous study on its effectiveness to filter the given input signal accurately with the least possible error, the conclusion obtained is as follows: RLS is the most effectual followed by NLMS and LMS. This is because the convergence rate for RLS is higher than that of LMS and also the fact that RLS uses the present and past values to predict its future values whereas LMS uses only the present ones. NLMS is a better and an improved version of LMS and hence has a higher convergence rate by altering step size. All these conclusions have been drafted based on the results by comparing various parameters like SNR, cross correlation and MSE.

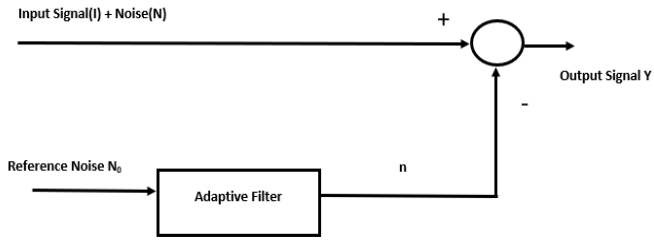

Fig.1. Basic block diagram of an ANC filter.

## II. ADAPTIVE NOISE CANCELLATION ALGORITHMS

Adaptive noise cancellation algorithms have a wide range of applications ranging from hearing aids to the large systems used in vehicles such as cars or airplanes. Although these applications are diverse, the integral part of the algorithms is very similar. Mainly, LMS, NLMS and RLS adaptive noise cancellation filters are used for speech enhancement.

An adaptive filter is time-varying since their parameters are continually changing in order to meet certain performance requirements where x(n) denotes the input signal, n is the noise added to the signal, n0 is the reference noise inputted, d (n) defines the reference or desired signal, error signal e (n) is the difference between the desired d (n) and filter output y (n). The error signal is used as a feedback to the adaptation algorithm in order to determine the appropriate updating of the coefficients of the filter [5].

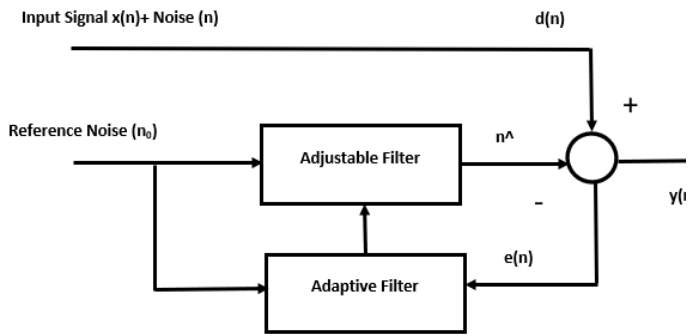

Fig.2. Block Diagram of an Adaptive Noise Cancellation System

As shown in Fig.2, an adaptive noise canceller (ANC) has two inputs- (i) primary and (ii) reference. The primary input receives a signal x from the signal source that is corrupted by the presence of noise n. The reference noise $n_0$ passes through a filter to produce an output nˆ that is a close estimate of the primary input noise. This noise estimate is then subtracted from the corrupted signal to produce an estimate of the output signal y (n) [6].

### A. Recurisve Least Square

The Recursive Least Squares (RLS) algorithm uses the method of least squares. Through the given approach, the difference between a desired and estimated signal is squared and summed in order to find a best fit. RLS uses a deterministic approach to adaptively find a best-fit filter for an adaptive filter system. For signals that are non stationary, this filter tracks the time variations but in the case of stationary signals, the convergence behavior of this filter is similar to Wiener filter which converges to the same optimal coefficients. This adaptive filter has complex computations and has high speed of convergence, minimum error at convergence, numerical stability and robustness [7]. The algorithm is defined by the following equation:

$$w(n+1) = w(n) + e(n).k(n) \qquad (1)$$

Where w(n) is the filter coefficients vector and k(n) is the gain vector. k(n) is defined by the following equation:

$$k(n) = \frac{p(n).u(n)}{m + u^T.p(n).u(n)} \qquad (2)$$

Where *m* is the forgetting factor and *p(n)* is the inverse correlation matrix of the input signal.
The value of the forgetting factor must always lie below 1 and before beginning any iteration, initialize the value w(0)=0.

The step size of an RLS filter which has a very good convergence rate can be between 0 and 1.

The RLS-type algorithms have a high convergence speed which is independent of the eigenvalue spread of the input correlation matrix.

### B. Least Mean Square

LMS algorithm is a filter in which the weights of the filter are obtained based on the least mean square error signal (difference between the input and the reference signals). The weights of the filter are updated in a manner that they converge to the actual filter weights. This type of adaptive filter has a simple design, good convergence rate (Convergence is the process of minimizing the error signal) and easy computations. It does not involve matrix operations as that of the RLS algorithm; as a result, the calculations are unproblematic. The working of the LMS filter is mainly divided into two subparts, (i) Filtering Process, wherein the output of the filter and the mean square error signal is m determined by comparing the output signal to the desired signal. (ii) Adaptive Process, wherein the weights of the filter are adjusted based on the obtained error signal [8]. The Algorithm can be determined using the equation given below:

$$w(n+1) = w(n) + \mu.e(n).u(n) \quad (3)$$

Where $\mu$ is the step size of the adaptive filter, w(n) is the filter coefficients vector, and u(n) is the filter input vector.
The output of the filter is given by the following equation:

$$s(n) = u^T(n) + w(n) \quad (4)$$

After the calculation of the output of the filter, the error signal is calculated after which the weights are readjusted using equation (4) and the output is calculated repeatedly till the error signal is minimized.
The maximum step size $\mu_{max}$ can be easily calculated from the input samples of the LMS filter. The range of $\mu$ provided by the inequality above is sufficient for the stability of the LMS algorithm but not necessary.

$$\mu \leq \frac{1}{\lambda max} \quad (5)$$

where $\lambda\, max$ denotes the maximum eigenvalue of the autocorrelation matrix R of the input vector x(n).
The above range does not necessarily satisfy the stability condition of the LMS algorithm. The convergence of the mean of w(n) towards w0 and also the convergence of the variance of the elements of w(n) to a few limited values is necessary for the convergence of the LMS algorithm. [18]

*C. Normalized Least Mean Square*

The main drawback of LMS algorithm is that it is sensitive to the scaling of input values. This makes it very hard to select a rate $\mu$ that guarantees stability of the algorithm. The Normalized least mean squares (NLMS) filter is a better variant of the LMS algorithm that solves this problem by normalizing the power of the input [9]. NLMS algorithm updates the coefficients of the adaptive filter by using the following equation,

$$w(n+1) = w(n) + \mu.e(n).\frac{u(n)}{\|u(n)\|^2} \quad (6)$$

$$w(n+1) = w(n) + \mu.e(n).u(n) \quad (7)$$

Where, $u(n) = \frac{u(n)}{\|u(n)\|^2}$

The need to derive NLMS algorithm is that, with the change in the input signal power with time the step size of two adjacent coefficients of the filter will also change, which affects the convergence rate. This causes reduction in convergence rate of weak signal and increment in the convergence rate of the strong signal. To overcome this problem, the step size has to be changed with respect to the input power. Therefore the step size parameter is said to be normalized. This algorithm may be a suitable alternative which normalizes the LMS step size with the power of the input signal.

Hence, it can be found that the step size of NLMS filter can vary from 0 to 2 thus providing maximum convergence.[18]

III. IMPLEMENTATION AND ANALYSIS

Implementation of adaptive noise cancellation using LMS, NLMS and RLS filters requires an input signal which is a speech signal in this case. Noise is added through the channel to our speech signal. This added noise creates problems by corrupting the original signal.

In this paper, to the audio signal additive white Gaussian noise and random noise have been added to corrupt the signal. Algorithms like RLS, LMS, NLMS have been incorporated to free the signal from the two noises that have been added. This noise removal process involves two inputs, one being the audio signal and the other being the reference noise signal.

To conduct this, a procedure needs to be followed to add the noise to the speech signal and then removing it through the algorithms mentioned above. Each algorithm has its own mathematical calculations that have to be followed. The general procedure has been elucidated below.

To de-noise the corrupted signal, the following steps have been followed:
Step 1: Record a speech signal using MATLAB.
Step 2: Add white noise or random noise to the recorded speech signal and plot their respective graphs.
Step 3: Choose between LMS, NLMS and RLS adaptive filter for active noise cancellation of the corrupted Speech signal.
Step 4: Vary the variables which are associated with the adaptive noise cancellation filters such as filter length and step size.
Step 5: Plot the denoised speech signal obtained as the output of the adaptive noise cancellation filter.
Step 6: calculate SNR (signal to noise ratio), Correlation coefficient and MSE (mean squared error).
Step 7: Repeat the same procedure for the other adaptive noise cancellation filters.
Step 8: Compare and tabulate the obtained results.
Thus, in this way, we can implement the LMS, NLMS and RLS adaptive noise cancellation algorithms

IV. SIMULATIONS, RESULTS AND DISCUSSIONS

The adaptive filter algorithms applied in this paper, LMS, NLMS and RLS, have been implemented and simulated using MATLAB. The comparison between the algorithms are done using directives such SNR, cross correlation and mean squared error. At first the speech signal is recorded, wherein the speaker says "Hello, Good morning. It's a beautiful day outside and the sun is shining brightly in the sky." Fig.3 shows the input speech with the recorded message, which is converted into a vector.

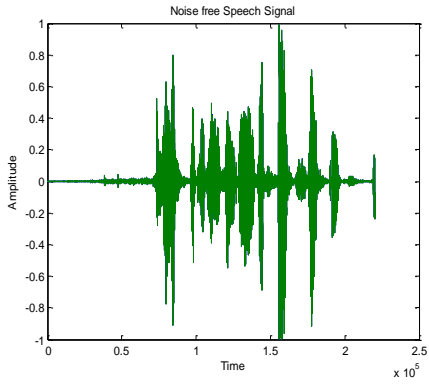

Fig.3. Input Speech signal

This speech signal is converted into a vector and two types of noises are added to it, White Gaussian noise and Random noise.

Now this input speech signal with noise is passed through the filter as one input, while the other input is the reference noise. The outputs of the three adaptive filters are compared on the basis of SNR, correlation coefficient and the mean squared error for varying filter length(order) N=5 and N=10 and varying step size µ=0.05 and µ=0.1. Higher the order of the filter higher is the accuracy but the cost of the hardware realization of the filter increases. The step size of the filter has to be within the range of 0 to 2 for the adaptive filter to be implemented in MATLAB. Table 3.1 shows the implementation of LMS, NLMS and RLS filters for the filter length N=5

TABLE.3.1 Comparative Results of NLMS, LMS and RLS Filter for Length N= 5.

| PARAMETERS | | WHITE NOISE | | | RANDOM NOISE | | |
|---|---|---|---|---|---|---|---|
| STEP SIZE | METRICS | NLMS | LMS | RLS | NLMS | LMS | RLS |
| 0.05 | SNR | 0.3321 | -0.0043 | 3.0563 | -0.0013 | -0.0409 | 2.0081 |
| | Correlation coefficient | 0.7374 | 0.6278 | 0.8619 | 0.6945 | 0.5831 | 0.7973 |
| | MSE | 0.0042 | 0.0227 | 0.0009 | 0.0203 | 0.0496 | 0.0089 |
| 0.10 | SNR | 0.1781 | -0.0069 | 1.3172 | -0.0973 | -0.2461 | 0.8163 |
| | Correlation coefficient | 0.7156 | 0.6948 | 0.7536 | 0.6513 | 0.6073 | 0.7291 |
| | MSE | 0.0219 | 0.2718 | 0.0021 | 0.0482 | 0.0716 | 0.0197 |
| 0.15 | SNR | 0.0962 | 0.0274 | 0.5179 | -0.2871 | -0.3879 | 0.3975 |
| | Correlation coefficient | 0.5993 | 0.5136 | 0.6971 | 0.5219 | 0.4988 | 0.6468 |
| | MSE | 0.0753 | 0.0842 | 0.0619 | 0.1385 | 0.1549 | 0.0954 |

From Table 3.1, it can be observed that as the step size increases from 0.05 to 0.1, the SNR of the output signal decreases for a given order of the filter. Another issue to be noted here is that the SNR of the output signal in both cases of random noise and white noise is quite low. The SNR of NLMS and LMS filter is negative stating that the noise is more prominent than signal. Though the SNR of RLS is positive, it is not exceptionally high. It can also be observed that the MSE increases with the decrease in SNR.

TABLE.3.2 Comparative Results of NLMS, LMS and RLS Filter for Length =10

| PARAMETERS | | WHITE NOISE | | | RANDOM NOISE | | |
|---|---|---|---|---|---|---|---|
| STEP SIZE | METRICS | NLMS | LMS | RLS | NLMS | LMS | RLS |
| 0.05 | SNR | 7.7563 | 4.5631 | 11.6547 | 4.2403 | 2.7135 | 7.1872 |
| | Correlation coefficient | 0.7982 | 0.7247 | 0.8721 | 0.7061 | 0.6912 | 0.8153 |
| | MSE | 0.0023 | 0.0972 | 0.0006 | 0.0035 | 0.0064 | 0.0008 |
| 0.10 | SNR | 5.4926 | 2.2479 | 8.7852 | 3.8766 | 1.8729 | 5.9253 |
| | Correlation coefficient | 0.8163 | 0.7758 | 0.9141 | 0.7880 | 0.7674 | 0.8631 |
| | MSE | 0.0051 | 0.0107 | 0.0009 | 0.0097 | 0.0843 | 0.0035 |
| 0.15 | SNR | 3.9981 | 2.5671 | 6.1293 | 2.5674 | 1.0041 | 4.9831 |
| | Correlation coefficient | 0.8563 | 0.8109 | 0.9301 | 0.8238 | 0.8019 | 0.8994 |
| | MSE | 0.0118 | 0.1197 | 0.0013 | 0.1307 | 0.1863 | 0.0154 |

From table 3.2, we can observe that the SNR has increased from that of table 3.1 as the order has increased which has further led to the decrease of the MSE. As a result, we increased the filter length to N=15 to prove the results.

TABLE.3.3 Comparative Results of NLMS, LMS and RLS Filter for Length = 15

| Filter length N=15 | | | | | | | |
|---|---|---|---|---|---|---|---|
| | | WHITE NOISE | | | RANDOM NOISE | | |
| STEP SIZE | PARAMETERS | NLMS | LMS | RLS | NLMS | LMS | RLS |
| 0.05 | SNR | 15.8729 | 13.7347 | 18.9638 | 12.7493 | 10.4892 | 15.9683 |
| | Correlation coefficient | 0.8497 | 0.8013 | 0.9016 | 0.8159 | 0.7631 | 0.8971 |
| | MSE | 0.0015 | 0.1342 | 0.0002 | 0.0213 | 0.1983 | 0.0005 |
| 0.10 | SNR | 12.0167 | 9.9981 | 14.9719 | 9.9786 | 8.1638 | 12.5289 |
| | Correlation coefficient | 0.8971 | 0.8579 | 0.9159 | 0.8758 | 0.8294 | 0.9026 |
| | MSE | 0.0215 | 0.1749 | 0.0006 | 0.0297 | 0.0357 | 0.0017 |
| 0.15 | SNR | 10.0317 | 7.6581 | 11.8972 | 8.6937 | 6.1036 | 9.9981 |
| | Correlation coefficient | 0.9104 | 0.8842 | 0.9206 | 0.8991 | 0.8569 | 0.9143 |
| | MSE | 0.0284 | 0.2135 | 0.0011 | 0.3107 | 0.4096 | 0.0138 |

From the table 3.3, it can be inferred that SNR has increased as compared to the previous table at N=10. It can be stated that white noise can be removed much easily than that of random noise when mixed with an input audio signal based on the various values of SNR that can be verified using the table. The following graphs shown in the Fig.4 to Fig.9 are the output graphs obtained after the noisy signal was passed through the various algorithms.

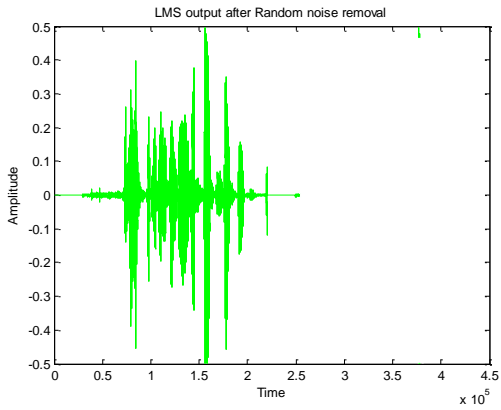

Fig.4. Output of LMS after Random Noise removal.

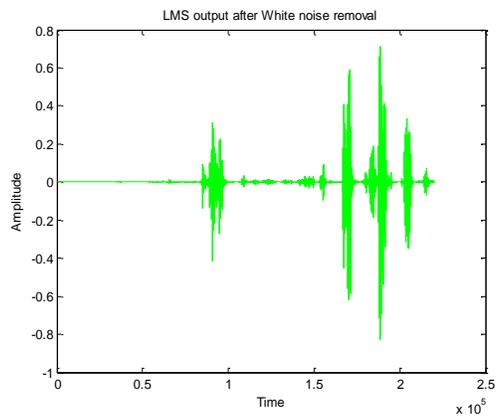

Fig.5. Output of LMS after white Noise removal.

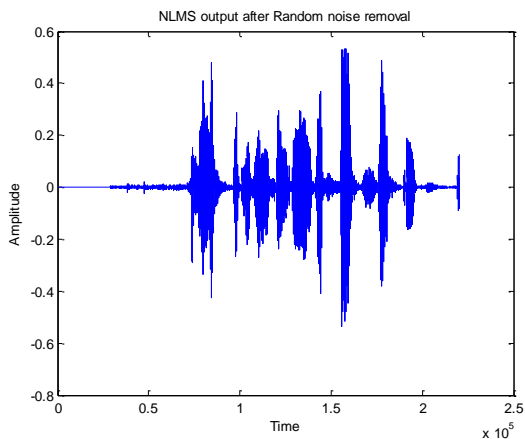

Fig.6. Output of NLMS after Random Noise removal.

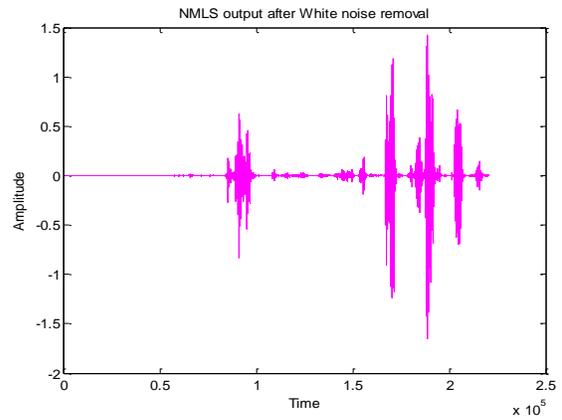

Fig.7. Output of NLMS after White Noise removal

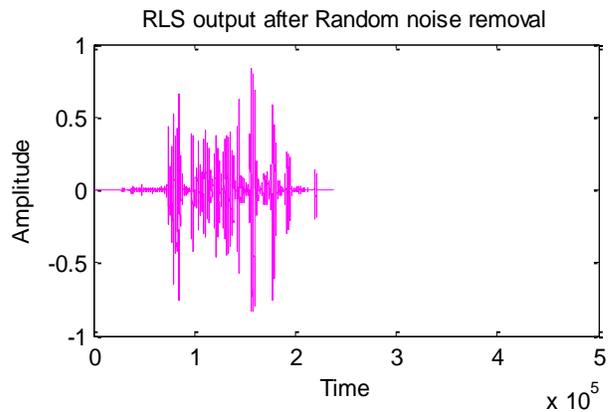

Fig.8. Output of RLS after Random Noise removal.

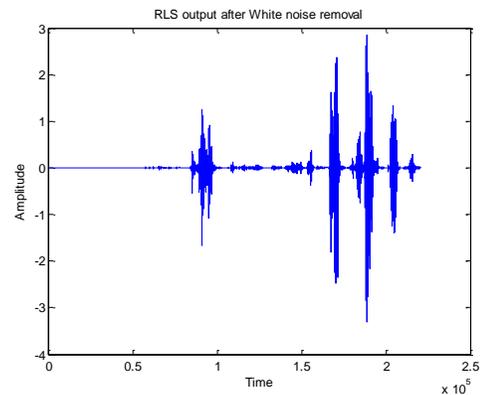

Fig.9. Output of RLS after White Noise removal.

From the given output graphs, it can be seen that RLS filter works best as its denoising capability is higher than that of LMS and NLMS as it has a higher convergence rate and produces an output which resembles the input completely, much better than that of the NLMS and LMS filters which do not completely cancel out the noise.

V. CONCLUSION

A comparative study on the effectiveness of the ANC filters when subjected to random and white noise has been done using parameters such as varying step size, SNR, correlation coefficient and MSE (Mean-squared error) has been implemented in this paper. Based on the results and graphs obtained, we can conclude that the RLS filter is the most efficient noise reduction filter followed by NLMS and LMS filters respectively. Signal reconstruction is best in the RLS filter due to a higher convergence rate of the filter followed by the NLMS and LMS filters respectively. Though the RLS output is much more favorable than the other two filters, it can be noted that RLS is a highly unstable filter as it undergoes lot of computations to provide the output. It can also be concluded that random noise cannot be removed as easily as white noise due to its unpredictable nature. Hence, RLS filter at length N=15 at step size (μ) of 0.05 proves to be the best filter as the SNR is the highest and the Mean squared error is the least.

## References

[1] Dixit, Shubra and Deepak Nagaria."LMS Adaptive Filter for Noise Cancellation: A Review" *International Journal Of Electrical and Computer Engineering* 7.5(2017):2520.

[2] Haykin, Simon S. *Adaptive filter theory*, Pearson Education India, 2008.

[3] Proakis JG, Manolakis DG. Digital signal processing 4th edition, 2006.

[4] Mendiratta, Arnav, and Devendra Jha, "Adaptive noise cancelling for audio signals using least mean square algorithm." *Electronics, Communication and Instrumentation (ICECI), International Conference on*. IEEE, 2014.

[5] Thenua, R. K., & Agarwal, S. K.," Simulation and performance analysis of adaptive filter in noise cancellation", *International Journal of Engineering Science and Technology*, 2010.

[6] Pandey, Deepak, and Sunder Raj Patel Ankit., "Real time active noise cancellation using adaptive filters following RLS and LMS algorithm",2016.

[7] Borisagar, Komal R., and G. R. Kulkarni. "Simulation and comparative analysis of LMS and RLS algorithms using real time speech input signal." *Global Journal of Research in Engineering*, 2010.

[8] Panda, S., & Mohanty, M. N.,"Performance analysis of LMS based algorithms used for impulsive noise cancellation, In *Circuit, Power and Computing Technologies (ICCPCT), International Conference*, IEEE, *2016.*

[9] Gupta, Priyanka, Mukesh Patidar, and Pragya Nema. "Performance analysis of speech enhancement using LMS, NLMSand UNANR algorithms." *Computer, Communication and Control (IC4), International Conference on*. IEEE, 2015.

[10] Sharma, Lalita, and Rajesh Mehra. "Adaptive Noise Cancellation using Modified Normalized Least Mean Square Algorithm."

[11] Ferdouse L, Akhter N, Nipa TH, Jaigirdar FT. Simulation and performance analysis of adaptive filtering algorithms in noise cancellation, 2011 Apr 11.

[12] Luo, X.D., Jia, Z.H. and Wang, Q., "A new variable step size LMS adaptive filtering algorithm. *Acta Electronic Sinica"*, 2006.

[13] Liu, Y., Xiao, M., & Tie, Y.," A Noise Reduction Method Based on LMS Adaptive Filter of Audio Signals", In *3rd International Conference on Multimedia Technology, ICMT*, 2006.

[14] Divya, Preeti Singh, and Rajesh Mehra. "Performance Analysis of LMS & NLMS Algorithms for Noise Cancellation." *International Journal of Scientific Research Engineering & Technology (IJSRET)* 2, no. 6,2013.

[15] Widrow, Bernard, John R. Glover, John M. McCool, John Kaunitz, Charles S. Williams, Robert H. Hearn, James R., Zeidler, JR Eugene Dong, and Robert C. Goodlin. "Adaptive noise cancelling: Principles and applications." *Proceeding of the IEEE* 63, 1975.

[16] Sridhar, Bharath, I. Akram Sheriff, KA Narayanan Kutty, and S. Sathish Kumar. "Comparison of cascaded LMS-RLS, LMS and RLS adaptive filters in Non-Stationary environments." In *Novel Algorithms and Techniques in Telecommunications and Networking*, Springer, Dordrecht, 2010.

[17] Vaseghi, Saeed V. *Advanced digital signal processing and noise reduction*. John Wiley & Sons, 2008.

[18] Sharma, R. A. G. H. A. V. E. N. D. R. A., and V. PREM Pyara. "A comparative analysis of mean square error adaptive filter algorithms for generation of modified scaling and wavelet function." *Int J Eng Sci Technol* 4.4 (2012): 1396-1401.69